\documentclass{article}

\usepackage{spconf,amsmath,graphicx}
\usepackage[hidelinks]{hyperref}
\usepackage[inline]{enumitem}
\usepackage{listings}
\usepackage{subfig}
\usepackage{cite}
\usepackage{booktabs}
\usepackage{multirow}
\usepackage{enumitem}
\usepackage{comment}
\usepackage{xcolor}
\usepackage{colortbl}

\definecolor{highlight}{HTML}{ade6dc}

\title{FRONTEND TOKEN ENHANCEMENT FOR TOKEN-BASED SPEECH RECOGNITION}

\name{Takanori Ashihara, Shota Horiguchi, Kohei Matsuura, Tsubasa Ochiai, and Marc Delcroix}
\address{NTT, Inc., Japan}
\begin{document}

\ninept

\maketitle
\begin{abstract}
Discretized representations of speech signals are efficient alternatives to continuous features for various speech applications, including automatic speech recognition (ASR) and speech language models.
However, these representations, such as semantic or phonetic tokens derived from clustering outputs of self-supervised learning (SSL) speech models, are susceptible to environmental noise, which can degrade backend task performance.
In this work, we introduce a frontend system that estimates clean speech tokens from noisy speech and evaluate it on an ASR backend using semantic tokens.
We consider four types of enhancement models based on their input/output domains: wave-to-wave, token-to-token, continuous SSL features-to-token, and wave-to-token.
These models are trained independently of ASR backends.
Experiments on the CHiME-4 dataset demonstrate that wave-to-token enhancement achieves the best performance among the frontends.
Moreover, it mostly outperforms the ASR system based on continuous SSL features.
\end{abstract}

\begin{keywords}
discrete speech representation, semantic token, automatic speech recognition, speech enhancement, noise robustness
\end{keywords}

\vspace{-0.1cm}
\section{Introduction}
\label{sec:intro}
\vspace{-0.1cm}
Discretized speech representations have recently emerged as a key research direction for building efficient speech processing systems.
For example, semantic or phonetic tokens can be obtained by applying $k$-means clustering to features derived from self-supervised learning (SSL) speech models.
These representations have been successfully applied to automatic speech recognition (ASR)~\cite{compara_token_system,token_challenge,token_asr_tts} and speech language models (SLMs)~\cite{slm_survey}.
The generated tokens typically undergo sequence-shortening techniques such as deduplication~\cite{dedup_s2st} and byte-pair encoding (BPE)~\cite{bpe}, thereby facilitating more efficient data transmission, model training, and decoding.
\par
Despite such progress, noise robustness of ASR systems using semantic tokens (hereafter, token ASR) has not been sufficiently investigated.
In real-world scenarios, recordings are frequently corrupted by environmental noise, which can severely degrade ASR performance.
Therefore, improving noise robustness of ASR is a fundamental issue for practical deployment.
Several benchmark challenges have been established to drive progress in this area~\cite{chime4,chime7}.
In these challenges, ASR systems using continuous-valued vectors (hereafter, continuous ASR), such as log mel-filterbank features (FBANK) or SSL-derived features, have achieved top performance when combined with a speech enhancement (SE) frontend~\cite{gss,tf_gridnet}.
Moreover, some studies have reported further improvements by jointly fine-tuning an SE frontend with ASR objectives~\cite{iris}.
\par
While a large body of research has addressed robustness in continuous ASR, the potential of frontend enhancement for robust token ASR remains largely unexplored.
More concretely, since token ASR takes discrete representations as input, it is unclear \textit{whether enhancement should be performed at the waveform level, as input to SSL models, or directly at the discrete token level}.
This gap underscores the need for investigating whether token ASR can achieve both efficiency and robustness under noisy conditions.
\par
In this paper, we introduce multiple frontend enhancement strategies and explore their effectiveness in improving the noise robustness of token ASR backends.
To enable a systematic comparison, we categorize enhancement frontends according to the input/output representations used to train them: wave-to-wave (W2W-E; conventional SE), token-to-token (T2T-E), vector-to-token (V2T-E; where the vector is a weighted-sum feature from an SSL model), and wave-to-token enhancements (W2T-E).
Since the enhancement frontends are trained independently of the ASR backend, this modularization allows the enhancement component to remain unchanged even when the ASR model is replaced or updated.
\par
The main contributions of this work are as follows:
\begin{itemize}[nosep, topsep=0pt,
  leftmargin=*,
  labelindent=10pt,
  labelsep=5pt,
  itemindent=0pt,
  align=parleft
]
\item We present the first comprehensive and systematic evaluation of diverse enhancement frontends to improve the noise robustness of token ASR systems, including the introduction of novel V2T-E and W2T-E approaches.
\item Experiments on the CHiME-4~\cite{chime4} dataset demonstrate that W2T-E, despite its simplicity, yields the best token ASR performance among all frontends, even surpassing continuous ASR using weighted-sum SSL features in most cases.
\item We show that ASR accuracy does not always correlate with token-level accuracy, measured using reference and enhanced tokens, indicating the limitations of this metric in predicting recognition accuracy.
\end{itemize}
We expect that this work will serve as a foundation for future research on frontend enhancement in token-based speech processing.

\vspace{-0.1cm}
\section{Related work}
\vspace{-0.1cm}
Several studies have proposed semantic tokens designed to remain invariant under different types of perturbations.
For example, the standard $k$-means quantizer was replaced with a multi-layer perceptron (MLP) quantizer trained on SSL features~\cite{token_enh_ctc}.
This approach improved augmentation invariance and achieved better performance on a textless speech-to-speech translation task~\cite{dedup_s2st,textless_s2st}, where tokens serve as the training targets.
Similarly, a quantizer trained on SSL features was proposed to preserve local information under various distortions while disentangling global attributes such as speaker identity~\cite{nast}.
This approach yielded strong results on zero-resource speech benchmarks~\cite{zero_resource}.
Both approaches can be categorized as V2T-E methods.
Other perturbation-invariant tokenizers have also been explored, for example to improve robustness against speaker variation~\cite{textless_s2st,contentvec,spin}.
Building on these efforts, we explore the applicability of noise-invariant semantic tokens as a frontend enhancement for token ASR.
Moreover, we conduct a comprehensive evaluation across different input/output domains, extending beyond V2T-E, which was considered in prior studies~\cite{token_enh_ctc,nast}.
This provides a broader perspective on input/output designs and on how enhancement strategies can be integrated into token ASR pipelines.

\vspace{-0.1cm}
\section{Frontend enhancement for token ASR}
\label{sec:met}
\vspace{-0.1cm}
\begin{figure}[t]
  \centering
  \includegraphics[width=\linewidth]{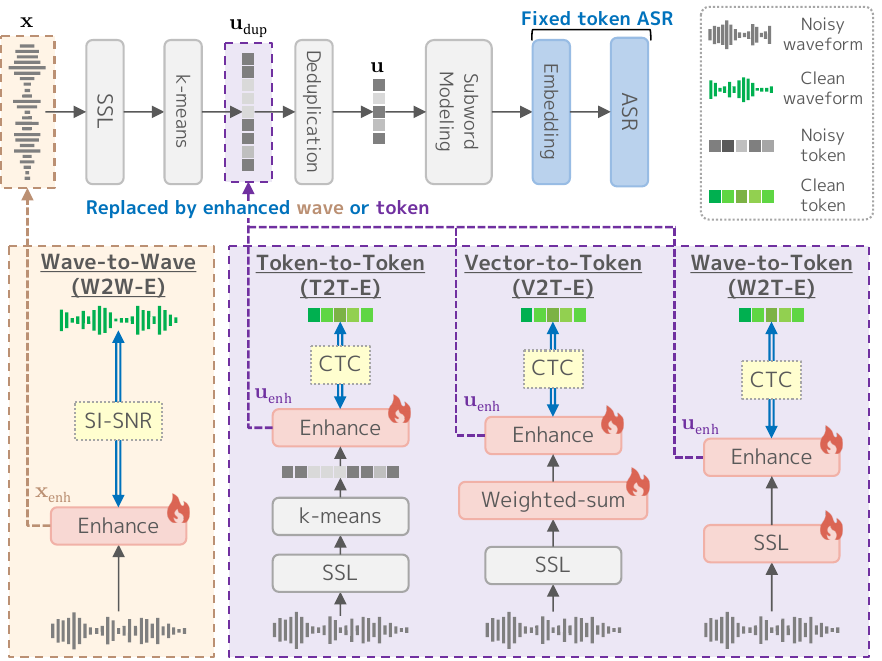}
  \vspace{-0.5cm}
  \caption{Schematic illustration of the token ASR backend (top) and the four categories of enhancement frontends (bottom) to improve backend noise robustness.}
  \label{fig:main}
\vspace{-0.3cm}
\end{figure}
This paper employs a semantic token ASR backend~\cite{compara_token_system}, which is shown in the upper part of Fig.~\ref{fig:main}.
Specifically, a speech signal $\mathbf{x}$ is converted into a token sequence $\mathbf{u}_\mathrm{dup}$ via SSL feature extraction followed by assigning $k$-means cluster labels.
The sequence $\mathbf{u}_\mathrm{dup}$ is deduplicated to obtain $\mathbf{u}$, then further compressed by subword modeling (e.g., BPE).
For token ASR training, a learnable embedding layer is incorporated into the input layer of the model to map the discrete tokens into the ASR model representation space.
\par
However, in noisy recordings, the tokens tend to deviate from their clean counterparts, leading to reduced ASR performance.
To address this, we investigate strategies for enhancing tokens to achieve noise-robust token ASR.
Conventionally, for continuous ASR, SE systems (referred to as W2W-E) are incorporated as a frontend to transform a noisy signal into an enhanced signal $\mathbf{x}_\mathrm{enh}$ (lower left part of Fig.~\ref{fig:main}).
For token ASR input, $\mathbf{x}_\mathrm{enh}$ is then passed through SSL feature extraction and tokenization to produce enhanced tokens.
\par
To comprehensively investigate modularized frontends, this study introduces \textit{token-level} enhancement schemes in addition to \textit{waveform-level} enhancements (i.e., W2W-E).
The schemes directly produce enhanced token sequences $\mathbf{u}_\mathrm{enh}$, which can serve as input to a trained token ASR, as illustrated in the lower right part of Fig.~\ref{fig:main}.
Since the output is discrete, the training objective can be either cross-entropy or connectionist temporal classification (CTC).
Based on our preliminary experiments, we employed CTC on deduplicated tokens as a training target, as it slightly outperformed cross-entropy on duplicated tokens.
\par
To systematically compare token-level enhancement frontends, we introduce three approaches, each defined by the training-time input representations---tokens, continuous vectors, or waveforms.
Compared to tokens, the latter two representations inherently offer richer information and thus have the potential for higher accuracy.
However, their higher dimensionality makes learning more challenging, requiring a more carefully designed optimization strategy.

\vspace{-0.1cm}
\subsection{Token-to-token enhancement (T2T-E)}
\label{sssec:T2T-E}
\vspace{-0.1cm}
The T2T-E frontend is a simple network that learns the mapping from the duplicated noisy tokens to $\mathbf{u}_\mathrm{enh}$.
This frontend employs a trainable embedding layer at the input, similar to token ASR.
The mapping network is motivated by recent work on SLMs, where several studies have highlighted their effectiveness for token-based SE~\cite{selm,llase}.
A similar architecture was also adopted in the baseline system of the discrete audio and speech benchmark~\cite{dasb}.

\vspace{-0.15cm}
\subsection{Vector-to-token enhancement (V2T-E)}
\label{sssec:V2T-E}
\vspace{-0.15cm}
The V2T-E frontend is inspired by \cite{token_enh_ctc}, which trains an MLP on SSL features extracted from a specific layer (i.e., the layer used for $k$-means clustering), using a CTC loss to build a perturbation-invariant tokenizer.
During MLP training, the parameters of the SSL model are kept fixed.
This process can be regarded as the distillation of quantization knowledge from the $k$-means model while incorporating noise invariance.
In our study, we explore the applicability as a token enhancement for token ASR and extend this strategy in two ways:
\begin{enumerate*}[label={\roman*)}]
\item replacing the tokenizer (i.e., MLP) with a more powerful model, and
\item using weighted-sum features.
\end{enumerate*}
\par
For the tokenization models, in addition to the original MLP, we evaluate E-Branchformer~\cite{ebranch} and a temporal convolutional network (TCN)~\cite{tcn}.
E-Branchformer enhances local aggregation and has been shown to outperform conventional Transformer/Conformer models in ASR tasks.\footnote{Our preliminary experiments also demonstrated a similar trend in the token-level enhancement task.}
The TCN consists of stacked 1-D dilated convolutional blocks and is employed as the separator network in Conv-TasNet~\cite{conv_tasnet}.
\par
As the input features to V2T-E, we adopt weighted-sum representations, with trainable weights, from all the layers of an SSL model.
This approach aims to capture the hierarchical nature of SSL representations, which distribute different speech characteristics (e.g., speaker identity and linguistic information) across layers, as in SUPERB~\cite{superb}.
Note that a previous study~\cite{streaming_token} also proposed training a token estimation model with such weighted-sum features.
However, it focused on streaming scenarios under clean conditions, whereas we target denoising and enhancement.

\vspace{-0.15cm}
\subsection{Wave-to-token enhancement (W2T-E)}
\label{sssec:W2T-E}
\vspace{-0.15cm}
The W2T-E frontend is similar to V2T-E but differs in its use of the SSL model, which is directly trained as a tokenizer taking the noisy speech signal as input to estimate $\mathbf{u}_\mathrm{enh}$.
We attach a single linear layer on top of the SSL model and fine-tune the entire network using a CTC loss, as is commonly done in the ASR system initialized with an SSL model~\cite{wav2vec2,hubert,wavlm}.
Note that, as in V2T-E, we also investigated more powerful classifiers in place of the linear layer, but the performance remained almost the same.
The W2T-E design has the highest training cost among the token-level enhancements because the SSL model itself is trained as part of the system.
However, since it does not produce any intermediate representations (i.e., tokens or continuous vectors), it results in the simplest overall system and the lowest inference cost among the considered frontends.

\vspace{-0.2cm}
\section{Experimental Setup}
\label{sec:expset}
\vspace{-0.1cm}
\subsection{Evaluation metrics and datasets}
\vspace{-0.1cm}
We compared enhancement frontends by evaluating ASR performance in terms of word error rate (WER).
Continuous ASR was evaluated with waveform-level enhancements (i.e., W2W-E), whereas token ASR was assessed with both waveform- and token-level enhancements (i.e., T2T-E, V2T-E, and W2T-E).
To directly evaluate enhancement outputs, we also used scale-invariant signal-to-noise ratio (SI-SNR) and unit edit distance (UED) for waveform- and token-level enhancements, respectively.
Here, UED is defined as the Levenshtein distance between the deduplicated reference (clean) and noisy/enhanced token sequences, normalized by the length of the deduplicated reference token sequence~\cite{token_enh_ctc}.
\par
All experiments were conducted on the CHiME-4 dataset~\cite{chime4}.
We followed the ESPnet setups to train the continuous ASR, token ASR, and enhancement models using the corresponding recipes.\footnote{\texttt{espnet/egs2/chime4/\{asr1,asr2,enh1\}}}
We report performance only on the single-channel datasets.\footnote{\texttt{\{dt05,et05\}\_\{simu,real\}\_isolated\_1ch\_track}}
In addition, the clean speech of the evaluation subset \texttt{et05\_simu} was used to report the upper-bound performance of the ASR system (hereafter referred to as \texttt{clean}).
When computing UEDs, we used the simulated datasets (i.e., \texttt{\{dt05,et05\}\_simu}).
For simplicity, we omit ``\texttt{05}'' from the dataset names hereafter.

\vspace{-0.1cm}
\subsection{Models and training}
\vspace{-0.1cm}
\textbf{Backend ASR system:}
We trained a \textit{continuous ASR model} as the baseline, along with a \textit{token ASR model}.
For both ASR models, we adopted joint CTC/attention-based encoder-decoder models (hereinafter AED).
These models consisted of 12 E-Branchformer encoder blocks and 6 Transformer decoder blocks, following the configuration provided in the \texttt{asr2} recipe of ESPnet.
We also used a CTC-only token ASR with the same encoder architecture to test whether the enhanced tokens are reusable across different ASR.
\par
The continuous ASR models used either 80-dimensional FBANK features or weighted-sum features of WavLM Large~\cite{wavlm} as input.
The token ASR took 2k BPE units as input, derived from 1k clusters obtained via $k$-means clustering on the hidden representations from the 21st layer of WavLM Large.
The embedding dimension for the token ASR was set to 512.
For convolutional subsampling in the encoder, we applied two layers of 2D convolution with ReLU (reducing the sequence length to one quarter) for the FBANK-based ASR model.
For the other ASR models, we incorporated two layers of 1D convolution with ReLU (keeping the sequence length unchanged).
Other configurations followed the default ESPnet setups.
\par
To establish a stronger baseline, we further fine-tuned the WavLM-based continuous ASR model, including WavLM, initialized with the parameters of the trained AED.
The entire model was fine-tuned jointly with a learning rate of 5e-5 using a reduce-on-plateau scheduler with early stopping (patience of 5).
Note that the fine-tuned WavLM model lost its downstream-agnostic property, as the output features could no longer be used with other ASR models.
\par
Furthermore, to compare with existing systems, we decoded our ASR systems using a 16-block Transformer LM trained according to the ESPnet recipe.
\par
\textbf{Frontend enhancement system:}
For W2W-E, we focused on two widely used SE models: Conv-TasNet~\cite{conv_tasnet} and TF-GridNet~\cite{tf_gridnet}.
To build these models, we followed the setups provided in ESPnet.
\par
For T2T-E (Section~\ref{sssec:T2T-E}), the mapping network employed four E-Branchformer blocks, identical to the ASR encoder described above except for a reduced feature dimension of 256.
For the embedding layer, we adopted the same setup as in token ASR, based on our preliminary experiment\footnote{We also tested initializing the embedding layer using the centroids of the $k$-means clusters, but the performance gap was not significant.} (total 9.20M params).
\par
For V2T-E (Section~\ref{sssec:V2T-E}) with MLP, we used two linear layers of 512 dimensions with LeakyReLU activations, a similar architecture to that in~\cite{token_enh_ctc} (total 1.30M params).
For V2T-E with TCN, the architecture was composed of three TCN blocks.
Each block comprised eight 1D convolutional layers with a kernel size of 3, a hidden feature dimension of 512, and a bottleneck dimension of 128 (total 3.95M params).
V2T-E with E-Branchformer used the same architecture as the T2T-E mapping network (total 9.08M params).
Note that we report V2T-E results only with weighted-sum features of all SSL layers, as these features generally improved performance over those from the single layer used for tokenization.
\par
For W2T-E (Section~\ref{sssec:W2T-E}), the model was trained with LayerDrop~\cite{layerdrop} at a rate of 0.1, freezing the convolutional feature encoder throughout training and freezing the Transformer encoder for the first 15k steps\footnote{These freezing techniques, in particular, improved both WER and UED, whereas SpecAugment-like masking~\cite{wav2vec2,hubert,wavlm} did not improve either metric.}; the number of trainable parameters was 311.74M.
\par
All token-level frontend models were trained for 30 epochs with a batch size of 16.
For the learning rate schedule, a warmup scheduler with 25k warmup steps was used, except that V2T-E with MLP and TCN employed a reduce-on-plateau scheduler.
The optimizer was Adam, with learning rates of 5e-3 for T2T-E, 1e-4 for V2T-E with MLP, 1e-3 for V2T-E with TCN, 5e-3 for V2T-E with E-Branchformer, and 1e-4 for W2T-E.

\vspace{-0.1cm}
\section{Results}
\label{sec:res}
\vspace{-0.1cm}

\begin{table}[t]
\caption{Evaluation results on simulated and real datasets of CHiME-4~\cite{chime4} with various frontends, reported in terms of WER (\%) and UED. The results were obtained without LMs.}
\vspace{-0.3cm}
\label{tab:main_result}
\centering
\setlength{\tabcolsep}{3pt}
\resizebox{\linewidth}{!}{%
\begin{tabular}{@{}ll|ccccc|cc@{}}
\toprule
&&\multicolumn{5}{c|}{WER}&\multicolumn{2}{c@{}}{UED}\\
 &  & \texttt{dt\_} & \texttt{dt\_} & \texttt{et\_} & \texttt{et\_} & &\texttt{dt\_} & \texttt{et\_}\\
ID & System & \texttt{simu} & \texttt{real} & \texttt{simu} & \texttt{real} & \texttt{clean} & \texttt{simu} & \texttt{simu}\\
\midrule
\midrule
\multicolumn{2}{@{}l|}{\textbf{Continuous ASR (AED)}}&&&&&&&\\
B1 & FBANK & 18.0 & 15.1 & 25.1 & 23.0 & 7.4 & \multicolumn{2}{c}{-} \\
B2 & WavLM (Weighted-sum) & 8.1 & 6.0 & 11.0 & 6.8 & \colorbox{highlight}{\textbf{1.5}} & \multicolumn{2}{c}{-} \\
B3 & \:+ W2W-E (Conv-TasNet) & 9.1 & 6.0 & 17.2 & 13.7 & \colorbox{highlight}{\textbf{1.5}} & \multicolumn{2}{c}{-} \\
B4 & \:+ W2W-E (TF-GridNet) & 5.9 & \colorbox{highlight}{\textbf{3.8}} & 11.2 & 8.2 & \colorbox{highlight}{\textbf{1.5}} & \multicolumn{2}{c}{-} \\ 
\midrule
\multicolumn{2}{@{}l|}{\textbf{Token ASR (AED)}}&&&&&&&\\
A1 & WavLM & 17.3 & 12.6 & 18.6 & 13.5 & 3.4 & 63.6 & 65.7 \\ \midrule
A2 & \:+ W2W-E (Conv-TasNet) & 12.9 & 9.5 & 21.6 & 19.3 & 3.4 & 46.3 & 55.3 \\
A3 & \:+ W2W-E (TF-GridNet) & 9.2 & 6.7 & 15.1 & 12.4 & 3.2 & 42.1 & 48.7 \\ \midrule
A4 & \:+ T2T-E & 17.0 & 12.1 & 18.6 & 9.4 & 3.5 & 37.4 & 39.3 \\ \midrule
A5 & \:+ V2T-E (MLP) & 11.5 & 8.8 & 14.7 & 10.0 & 3.5 & 34.6 & 37.1 \\
A6 & \:+ V2T-E (TCN) & 10.4 & 8.0 & 13.7 & 9.4 & 3.6 & 32.0 & 34.5 \\
A7 & \:+ V2T-E (E-Branchformer) & 9.8 & 7.7 & 13.6 & 8.9 & 3.3 & 30.8 & 33.5 \\ \midrule
A8 & \:+ W2T-E & \colorbox{highlight}{\textbf{5.6}} & 4.5 & \colorbox{highlight}{\textbf{8.2}} & \colorbox{highlight}{\textbf{6.5}} & 3.4 & \colorbox{highlight}{\textbf{27.2}} & \colorbox{highlight}{\textbf{29.2}} \\
\midrule
\multicolumn{2}{@{}l|}{\textbf{Token ASR (CTC-only)}}&&&&&&& \\
C1 & WavLM & 21.9 & 16.2 & 23.6 & 17.3 & 5.1 & 63.6 & 65.7 \\
C2 & \:+ W2T-E & 6.6 & 6.1 & 9.9 & 8.2 & 4.6 & \colorbox{highlight}{\textbf{27.2}} & \colorbox{highlight}{\textbf{29.2}} \\
\bottomrule
\vspace{-0.7cm}
\end{tabular}%
}
\end{table}

\subsection{Performance comparison}
\vspace{-0.1cm}
Table~\ref{tab:main_result} summarizes the evaluation results.
For the baseline continuous ASR models, applying a conventional SE frontend (B3, B4) improved performance in some cases, but artifacts introduced by the SE systems could have adversely affected recognition, as shown in a previous study~\cite{se_artifact}.
Here, Conv-TasNet and TF-GridNet achieved SI-SNRs of 11.38 dB and 13.17 dB on \texttt{et\_simu}, respectively.
\par
For token ASR, the AED model without enhancement (A1) outperformed FBANK-based continuous ASR (B1) but underperformed WavLM-based continuous ASR (B2), consistent with previously reported results~\cite{compara_token_system}.
This trend indicates the importance of pre-trained knowledge provided by SSL models.
We also confirm that token ASR (A1) is sensitive to noise, similar to continuous ASRs (B1, B2), by comparing WERs between \texttt{et\_simu} and \texttt{clean}.
\par
For token ASR with W2W-E (A2, A3), Conv-TasNet sometimes increased WERs, whereas TF‑GridNet consistently reduced them.
This result suggests that achieving a certain level of SI-SNR with W2W-E consistently improves the accuracy of token ASR, which, interestingly, contrasts with the trend of continuous ASR (B3, B4) as described above.
\par
For T2T-E (A4), the frontend achieved better UEDs but in most cases worse WERs than TF-GridNet-based W2W-E (A3) (see Section~\ref{sec:anal1} for the detailed analyses).
\par
For V2T-E, it mostly outperformed T2T-E in both WER and UED even with the MLP-based model (A5), and more powerful architectures yielded additional improvements (A6, A7).
Notably, TCN (A6) showed comparable performance to E-Branchformer (A7) despite having fewer trainable parameters (3.95M vs. 9.08M), highlighting its parameter efficiency.
\par
Finally, W2T-E (A8) achieved the best overall performance under noisy conditions in both WER and UED.
Except for \texttt{dt\_real}, its WERs outperformed WavLM-based continuous ASRs (B2, B4) while also reducing computational cost due to shorter input sequences.
Specifically, W2T-E with BPE reduced the average token sequence length by 68.8\% and 68.2\% for \texttt{et\_real} and \texttt{et\_simu}, respectively, relative to the output frames of WavLM.
Additionally, the enhancement frontends, including W2T-E, exhibited no noticeable degradation on \texttt{clean}, indicating that they can reliably substitute for a $k$-means-based tokenizer.
Notably, W2T-E requires only a single additional linear layer, which keeps its computational overhead minimal during inference.
These results suggest that fully exploiting the capacity of SSL models requires more than simply adding a fixed tokenizer module (e.g., $k$-means tokenization).
\par
Furthermore, to verify the modularity of frontends, we evaluated our proposed W2T-E with the CTC-only token ASR backend (C1, C2).
The results of the C2 system confirm that it achieved substantial and consistent improvements, even when using a different backend.
\par
To clarify the position of our approach within the CHiME-4 single-channel sets, we compared our LM-based results with those of an existing system also using WavLM Large, called IRIS~\cite{iris}.
IRIS is an end-to-end optimized system that combines Conv-TasNet and an AED model on frozen WavLM features, achieving state-of-the-art performance.
As shown in Table~\ref{tab:lm_result}, our token ASR with W2T-E (A8) exhibited comparable performance to IRIS (E1), while benefiting from token ASR’s efficiency.
Notably, B5, obtained by continuing joint SSL-ASR fine-tuning of B2, yielded the best overall performance.
This suggests that jointly optimizing token-level enhancement, together with token ASR, might further reduce WER.
However, this is beyond the scope of this paper since such enhancements would be unsuitable for use with arbitrary token ASR systems.

\begin{table}[t]
\caption{WERs (\%) on single-channel datasets compared with an existing system. All systems were decoded with LMs. The results for the existing system are taken from the original paper~\cite{iris}.}
\vspace{-0.3cm}
\label{tab:lm_result}
\centering
\resizebox{\linewidth}{!}{%
\begin{tabular}{@{}ll|ccccc@{}}
\toprule
 & & \texttt{dt\_} & \texttt{dt\_} & \texttt{et\_} & \texttt{et\_} & \\
ID & System & \texttt{simu} & \texttt{real} & \texttt{simu} & \texttt{real} & \texttt{clean}\\
\midrule
\midrule
E1 & IRIS~\cite{iris} & 3.2 & \textbf{2.0} & 6.1 & 3.9 & - \\
\midrule
B2 & WavLM (Weighted-sum) & 5.4 & 3.7 & 8.0 & 4.4 & \textbf{0.8} \\
B4 & \:+ W2W-E (TF-GridNet)& 4.1 & 2.3 & 8.2 & 5.5 & \textbf{0.8} \\
B5 & \:+ Joint FT. WavLM & \textbf{3.1} & \textbf{2.0} & \textbf{5.6} & \textbf{3.5} & 1.0 \\
\midrule
A8 & WavLM (Token) + W2T-E & 3.2 & 2.3 & 6.1 & 4.0 & 1.6 \\
\bottomrule
\end{tabular}%
}
\vspace{-0.2cm}
\end{table}

\vspace{-0.15cm}
\subsection{Relationship between UED and WER}
\label{sec:anal1}
\vspace{-0.15cm}
\begin{figure}[t]
  \centering
  \includegraphics[width=\linewidth]{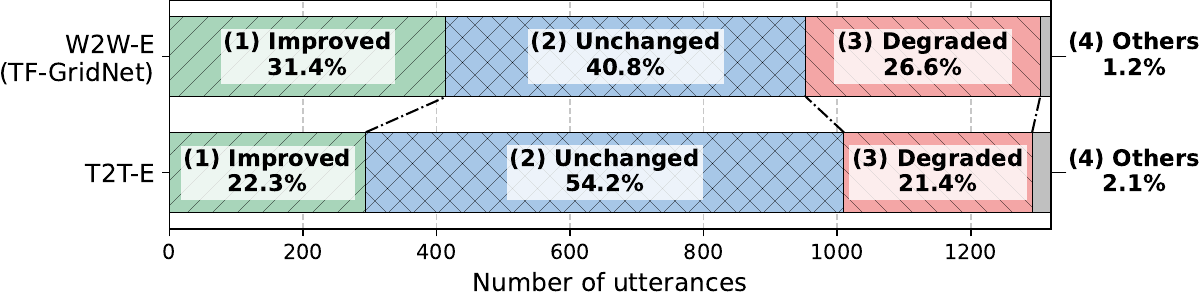}
  \vspace{-0.5cm}
  \caption{Utterance counts by UED-WER change group (\texttt{et\_simu}).}
  \vspace{-0.0cm}
  \label{fig:anal1}
\end{figure}
Based on the A3 and A4 results in Table~\ref{tab:main_result}, we conducted an in-depth analysis of the lack of correlation between UED and WER by calculating both metrics on a per-utterance basis.
In Fig.~\ref{fig:anal1}, utterances are grouped into four categories based on the changes in UED and WER observed in the enhanced tokens, compared with the original \texttt{et\_simu} noisy tokens:
\begin{enumerate*}[label=(\arabic*)]
\item both UED and WER improved,
\item UED improved while WER remained unchanged,
\item UED improved while WER degraded, and
\item UED was unchanged or degraded (``Others'').
\end{enumerate*}
Figure~\ref{fig:anal1} demonstrates that T2T-E had substantially more utterances with unchanged WER than TF-GridNet-based W2W-E, even though it achieved better average UED.
This observation suggests that token ASR exhibits a degree of robustness to noise-induced shifts in token sequences, at least to the extent covered by T2T-E.
In summary, while UED is a useful metric for assessing the robustness of tokenizers to noise, it should be interpreted with caution, as the inherent capacity of token-based backends to handle variability also plays a critical role.

\vspace{-0.15cm}
\subsection{Impact of SSL model depth on W2T-E}
\label{sec:anal2}
\vspace{-0.15cm}
\begin{figure}[t]
  \centering
  \includegraphics[width=\linewidth]{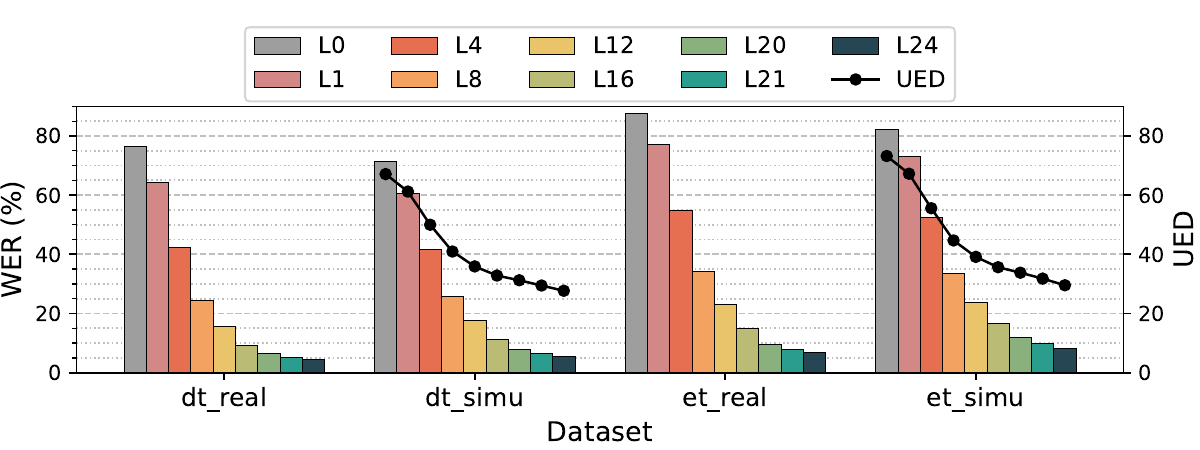}
  \vspace{-0.5cm}
  \caption{WERs (bars) and UEDs (black lines) as a function of SSL model depth in W2T-E. L0 denotes the output of the convolutional feature encoder of WavLM Large.}
  \vspace{-0.2cm}
  \label{fig:anal2}
\end{figure}
For the ablation study, we further investigated the effect of varying the depth of the SSL in W2T-E.
In training W2T-E, we examined whether attaching the CTC linear layer to an intermediate rather than the final WavLM layer could achieve comparable accuracy, since tokenization was originally based on the 21st layer.
If so, such a design would yield a more computationally efficient tokenizer.
To ensure a fair comparison, LayerDrop was not applied in this experiment, as it is not applicable to the zero-layer condition and its performance gain was marginal.
Figure~\ref{fig:anal2} shows WERs and UEDs for each dataset as a function of the W2T-E model depth.
The results reveal that a reduction in the number of layers is not feasible, as reducing the layers from 24 to 21 caused relative degradations of 16.2\% and 19.3\% in WER on \texttt{et\_real} and \texttt{et\_simu}, respectively.
One possible explanation is that, in addition to encoding phoneme or linguistic representations, ensuring robustness to noise may also require additional layers.

\vspace{-0.2cm}
\section{Conclusion}
\vspace{-0.2cm}
This paper introduced frontend enhancements for token ASR and explored the effectiveness of noise robustness.
Experimental results show that W2T-E offers substantial improvement for token ASR under noisy conditions, outperforming continuous-vector-based ASR, while requiring only an additional lightweight classifier.
As future work, we plan to extend the evaluation to other backend systems, i.e., speech tasks beyond ASR, particularly by using SLMs to further examine backend-agnostic characteristics.

\clearpage
\bibliographystyle{IEEEbib}
\bibliography{refs}

\end{document}